\documentstyle[prd,aps,psfig,twocolumn,floats]{revtex}

\begin{document}

\newcommand\D{\widehat{D}}
\newcommand\order[1] { ${{\cal O}\! \left( #1 \right)}$ }
\newcommand{\eq}[1]{Eq.\ (\ref{#1})}
\newcommand{\fig}[1]{Fig.\ \ref{#1}}
\newcommand{\lmax}{{\sf {L}}}
\newcommand{\mmax}{{\sf {M}}}
\newcommand{\ie}{{\em i.e.}}


\twocolumn[\hsize\textwidth\columnwidth\hsize\csname@twocolumnfalse\endcsname

\title{ Fast  Convolution on the Sphere}
\author{Benjamin D.~Wandelt$^{1,2}$ and Krzysztof M.~G\'{o}rski$^{3,4}$}
\address{${}^{1}$Department of Physics, Princeton University,
Princeton, NJ 08544, USA\\
${}^{2}$Theoretical Astrophysics Center, Copenhagen, Denmark\\
${}^{3}$European Southern Observatory, Garching bei M\"{u}nchen, Germany\\
${}^{4}$Warsaw University Observatory, Warsaw, Poland}
\date{\today}
\maketitle
\begin{abstract}
We propose fast, exact and efficient algorithms for the
convolution of two arbitrary functions on the sphere which speed
up computations by a factor \order{\sqrt{N}} compared to present
methods where $N$ is the
number of pixels. No simplifying assumptions are made other than
bandlimitation. This reduces typical computation times for
convolving the full sky with the asymmetric beam pattern of a
megapixel Cosmic Microwave Background (CMB) mission from months to
minutes. Our methods enable realistic simulation and careful
analysis of data from such missions, taking into account the
effects of asymmetric ``point spread functions'' and far side
lobes of the physical beam. While motivated by CMB studies, our
methods are general and hence applicable to the convolution or
filtering of any scalar field on the sphere with an arbitrary,
asymmetric kernel. We show in an appendix that the
same ideas can be applied to the inverse problems of map-making
and beam reconstruction by similarly accelerating the {\em transpose
convolution} which is needed for the iterative solution of the 
normal equations.
\end{abstract}
\hspace{1.5cm}
\maketitle
]

\section{Introduction}
A major near-term objective in the field of Cosmology today is to
gain a detailed measurement and statistical understanding of the
anisotropies of the cosmic microwave background (CMB). While the
theory of primary CMB anisotropy is well-developed (see
\cite{review} for a review) and we are  facing a veritable flood
of data from a new generation of instruments and missions,
perhaps the single most limiting factor for interpreting these
data is the exorbitant computational cost involved in realistic
mission simulation and careful analysis of the data
products\cite{borrill,BCJK}.

Important and computationally expensive tasks for both simulation
and analysis of microwave data are to simulate and to correct for
the systematic errors due to imperfections of realistic microwave
telescopes, such as beam asymmetries and far side lobes. The
effect of an asymmetric ``point spread function'' is to distort
the shapes of the detected anisotropies. What makes far side
lobes an important issue is the fact that the CMB anisotropy
signal has an amplitude of one in $10^5$ relative to the $2.7\,$K
background. In regions of low galactic latitude, foregrounds from
galactic synchrotron radiation and dust emission are expected to
exceed this signal by many orders of magnitude over a wide range
of frequencies \cite{haslam,SFD98}. Even though CMB experiments
will obviously  not target these regions to obtain measurements
of the background  anisotropy, the large amplitudes of these
galactic sources may induce  systematic errors even when
``looking'' in directions far away from the galactic plane if the
instrument allows diffraction of stray light into the detectors.
Solar system bodies, including the earth, are other possible
sources of stray light.

To assess these problems and formulate solutions we must be able
to compute the detector response at every pointing of the
telescope. The inputs are a physical model of the ``beam'' over
$4\pi$ steradian and a model of the ``sky'' containing both
simulated signal as well as foreground sources  possibly
including ground emission. Note that in the general case not just
the direction of the pointing is important but also the
orientation of the beam about the pointing axis. The detector
response is then the solution to a quadrature problem at each
orientation.

Analysis methods of  CMB data have neglected this difficulty by
assuming azimuthal symmetry of the beam which greatly simplifies
the calculation \cite{WS95,K99,Wu00}. Simulation work which did
include an asymmetric beam and far side lobes using pixel based
methods \cite{dM98,D98,B99} ran up against computational
challenges for angular scales smaller than one degree, running for
hundreds of hours even with optimized adaptive mesh algorithms.
Such algorithms are clearly inadequate for modern high resolution
experiments which achieve resolutions of a few minutes of arc.

In this paper we describe a numerical method which greatly
accelerates the computations which are necessary to correctly
account for realistic beam profiles in  simulation and  analysis
of directional data on the sphere. This is achieved by rewriting
the problem in such a way that we can take advantage of the
Cooley-Tukey Fast Fourier Transform (FFT) algorithm.

The following section of this paper defines the general problem in
terms of rotations of the beam with respect to the sky. We then
introduce a geometrically motivated split of the rotation
operator in section three. This enables us, in section four, to
derive the general solution for the detector response for {\em
all} possible relative orientations of the beam and the sky
within a given section on the sphere. Section five then discusses
the solution and derives special cases from it, amongst others the
well-known algorithm for convolution with azimuthally symmetric
kernels. We conclude in section six with a timing example. An appendix
applies the same 
ideas to accelerating the computation of the transverse
convolution, an operation which becomes important in the inverse
problem of map estimation.

While we are motivated by the goal of achieving and interpreting
precision measurements of the anisotropies of the cosmic
microwave background, the methods we present are general and
apply to the convolution or filtering of any scalar field on the
sphere with an arbitrary, asymmetric but constant kernel. We
generalise our methods to tensor fields on the sphere in
reference \cite{CMFAWG}.

\section{Statement of the Problem}
\label{definitions}
Consider two bandlimited functions on the sphere $b(\vec{\gamma})$ and
$s(\vec{\gamma})$. For definiteness and to aid the imagination
we will refer to them in the following as the {\em beam}
and the {\em sky}, respectively, but they could be completely general
bandlimited functions
--- in particular neither of them is constrained to be positive
definite or even real.

The task is to compute the scalar product of the beam and the sky at
a set of beam orientations.
To describe these orientations, we use the Euler
angles $\Phi_1,\Theta$ and
$\Phi_2$\footnote{Our Euler angle convention refers to active right
handed rotations of a physical body in a fixed coordinate system. The
coordinate axes stay in place under all rotations and the object
rotates around the $z$, $y$ and $z$ axes by $\Phi_1$, $\Theta$ and
$\Phi_2$, respectively, according to the right handed screw rule.}.
The convolved signal for each beam orientation $(\Phi_1,\Theta,\Phi_2)$
can then be written as
\begin{equation}
T(\Phi_2,\Theta,\Phi_1)= \int d\Omega_{\vec{\gamma}}\,
\left\lbrack\D(\Phi_2,\Theta,\Phi_1) b\right\rbrack\!\!(\vec{\gamma})^\ast s(\vec{\gamma}).
\label{eq:start}
\end{equation}
Here the integration is over all solid angles, $\D$
is the operator of finite rotations such that $\D  b$ is the
rotated beam, and the asterisk denotes complex conjugation.

If $(\Phi_1,\Theta,\Phi_2)$  can be written as a continuous
function of a parameter $t\in\left\lbrack0,T\right\rbrack$, say, then we call the ordered
set of tuples $(\Phi_1(t),\Theta(t),\Phi_2(t))$ a {\em scan
path}. Note that \eq{eq:start} assumes that time varying signals in
the sky vary either on time scales much longer than the duration of
the scan or much smaller than the integration time per sample. In the
context of CMB missions this is a good approximation with the
exceptions of  planets (for long
duration missions), time varying point sources, and  atmospheric foregrounds. Of these only
atmospheric foregrounds present a problem for the convolution, because they are
extended -  convolution with a point source is a simple operation in
position space (the pixel basis)
and can be computed separately. Linearity allows us to then add
the results of the $r\pi$ convolution of extended sources to the point
source convolution.

In the most general case, the bandlimits (see
\eq{eq:harmonic_expansion} below for a  definition)
of the beam and the sky are $\lmax_b$ and
$\lmax_s$, respectively. Define $\lmax\equiv {\rm
min}(\lmax_b,\lmax_s)$. Note that we actually  only need one
of $s$, $b$ to be bandlimited as long as the multipoles of the other
are bounded as $l\rightarrow\infty$. Then the numerical evaluation of  the integral in
\eq{eq:start}
takes \order{\lmax^2} operations for each tuple
$(\Phi_1,\Theta,\Phi_2)$. These integrals need to be evaluated for a
grid of beam locations that has to contain at least \order{\lmax^3}
grid points to allow subsequent interpolation at arbitrary locations.
Therefore the total computational cost for evaluating the convolution
using \eq{eq:start} scales as \order{\lmax^{{5}}}.

\section{Factorizing the rotation}
\label{sec:factor}
\begin{figure}[t]
\centerline{\psfig{file=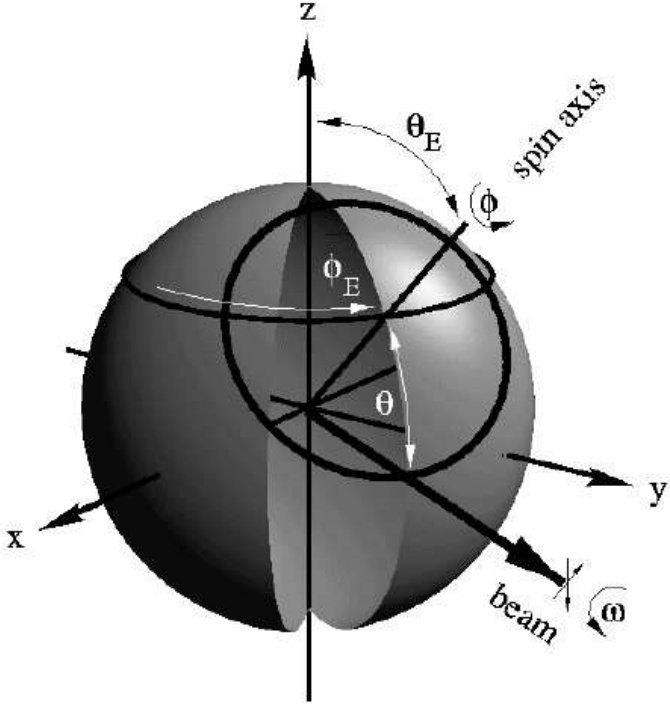,width=.5\textwidth}}
\caption{ Our coordinate system for efficient  convolutions. The beam
is shown at the position corresponding to $\theta=35^\circ$,
$\theta_E=50^\circ$, $\phi_E=60^\circ$, $\phi=0^\circ$ and
$\omega=0^\circ$. The cross-hairs on the beam mark its
orientation, here shown for $\omega=0$. In the null position ($\theta=\theta_E=\phi_E=\phi=\omega=0$) the beam is aligned with
the $z$-axis, the vertical cross hair pointing along
increasing $x$ and the horizontal cross hair pointing along increasing $y$.}
\label{fig:angles}
\end{figure}

It is possible to simplify the evaluation of \eq{eq:start}
significantly by factorizing the rotation into two auxiliary
rotations such as
\begin{equation}
\D(\Phi_2,\Theta,\Phi_1)\equiv\D(\phi_E,\theta_E,0)
\D(\phi,\theta,\omega).
\label{eq:rotsplit}
\end{equation}
We will define the various angles and motivate this split in the
following. Figure \ref{fig:angles} is intended to illustrate this
discussion.

To introduce these coordinates let us first consider {\em basic scan paths}. Imagine a scan path where
the beam sweeps
over the sky by scanning on rings of  angular radius
$\theta\in[0,\pi/2)$. The centers of these  scanning circles
lie on a ring of constant latitude, a polar angle
$\theta_E\in[0,\pi)$ away from the
north pole.
The angle $\phi_E\in[0,2\pi)$  selects a given scanning
circle and is defined as the longitude of its center, while
$\phi\in[0,2\pi)$
measures the angle along each scanning ring defined as increasing in a
right-handed way about the outward normal at the center, starting from zero at the
southernmost point on the ring.
Hence, for such a basic scan path we can  write the convolution as
set of scalar products  $T(\phi_E,\phi)$.
The angles $\theta$ and $\theta_E$ are thought of as  parameters
which fixed to define the scanning geometry.

As a generalization of basic scan paths, we allow as a further degree
of freedom an additional right handed rotation of the beam about its
outward axis by an angle $\omega\in[0,\pi/2)$.

Now we can see geometrically that all beam orientations on generalized
basic scan paths can be arrived at by successively applying the two
rotations in \eq{eq:rotsplit}. Define as the null position of the beam
when it is oriented along the $z$-axis
$\theta=\theta_E=\phi_E=\phi=\omega=0$.  Starting from the null
position, acting on it with
$\D(\phi,\theta,\omega)$ rotates it about its axis by
$\omega$ and moves it out onto a ring with opening angle
$\theta$ at an azimuthal angle $\phi$.  Then acting with
$\D(\phi_E,\theta_E,0)$ moves the beam into position.

Using the factorization \eq{eq:rotsplit}  we can re-write \eq{eq:start} as
\begin{eqnarray}
T(\phi_E,\phi,\omega)=\int d\Omega_{\vec{\gamma}}
\left\lbrack\D(\phi_E,\theta_E,0)\D(\phi,\theta,\omega)b\right\rbrack\!\!(\vec{\gamma})^\ast
s(\vec{\gamma}),\nonumber\\
(\theta,~\theta_E~{\rm fixed}).\label{eq:totalconv}
\end{eqnarray}

The function $T(\phi_E,\phi,\omega)$  contains all possible
integrals for a given scanning geometry. In fact, for the special case
$\theta=\theta_E=\pi/2$, these angles parameterize all possible
orientations of the beam on the sky, \ie~$(\phi_E,\phi,\omega)$
parameterize the group of rotations in
three dimensions. It is well known that in this case these coordinates
cover SO(3) twice, but this can be easily remedied by
restricting the range of one of the angles to half its range. We defer
removing this redundancy until the end of our calculation.

\section{Solution}
To exploit the form of \eq{eq:totalconv}, it is expedient to represent the functions $s$
and $b$ as well as the rotation operators  in the spherical harmonic
basis. A bandlimited function $f(\vec{\gamma})$
can be expanded in
spherical harmonics  as
\begin{equation}
f(\vec{\gamma})=\sum_{l=0}^{l=\lmax_f} \sum_{m=-l}^{m=l} f_{lm} Y_{lm}(\vec{\gamma}),
\label{eq:harmonic_expansion}
\end{equation}
where $\vec{\gamma}$ denotes a unit vector.
For practical applications the bandlimit $\lmax_f$ is chosen such
that higher terms contribute insignificantly.  We use the
notation where all quantities carrying both an $l$ and an $m$ index
vanish for $m>l$. This saves having to write explicit limits for
sums over azimuthal quantum numbers.

Invariance of the scalar product under a change of basis
then allows us to re-write \eq{eq:totalconv} as
$$
T(\phi_E,\phi,\omega)= \hspace{7.5cm}
$$
\begin{equation}
 \sum_{lm} s_{lm}
\left\lbrack\D(\phi_E,\theta_E,0) \D(\phi,\theta,\omega)
b\right\rbrack_{lm}^\ast =
\label{eq:derivation}
\end{equation}
$$
 \sum_{lmMM'}
s_{lm}
D^{l\;\ast}_{mM}(\phi_E,\theta_E,0)
D^{l\;\ast}_{MM'}(\phi,\theta,\omega)
b^{\;\ast}_{lM'}.
$$

A simple explicit expression for the matrix elements
$D^l_{m m'}(\phi_2,\theta,\phi_1)$ can be
given. One can define a real function $d^l_{m m'}(\theta)$ such that
\begin{equation}
D^l_{m m'}(\phi_2,\theta,\phi_1)=  e^{-im\phi_2}d^l_{m m'}(\theta)e^{-im'\phi_1}
\label{eq:D}
\end{equation}
Thus the dependence of $D$ on the Euler angles $\phi_1$ and $\phi_2$ is only in
terms of complex exponentials. While explicit formulas for the
$d$-functions exist \cite{BrinkSatchler}, they are more conveniently  computed
numerically using their recursion properties \cite{Risbo}.

Substituting into \eq{eq:derivation} and defining the
three--dimensional Fourier transform of $T(\phi_E,\phi,\omega)$ as
$$
T_{m\,m'\,m''}= \hspace{7.5cm}
$$
\begin{equation}
\frac1{(2\pi)^3}\int_0^{2\pi} d\phi_E d\phi d\omega \;
T(\phi_E,\phi,\omega) \, e^{-im\phi_E -im'\phi-im''\omega}.
\label{eq:FT}
\end{equation}
we obtain
\begin{equation}
 T_{m\,m'\,m''}=\sum_{l}
s_{lm} d^{l}_{mM}(\theta_E) d^{l}_{MM'}(\theta) b^{\;\ast}_{lM'}.
\label{eq:result}
\end{equation}

This equation is the main result of this paper, in effect generalising
fast 2D Fourier transfrom convolution from the plane to the sphere. Its
properties and specialisations will be discussed in the next section. Here
we give a geometrical interpretation. We have arrived at
\eq{eq:result} by writing convolution problems in such a way that
the results are fields on 3-tori instead of subsets of the
3-sphere, which is the group manifold of rotations in 3
dimensions. Convolutions over azimuthally symmetric and connected
sections of the 2-sphere (such as polar caps or annuli) can be
parameterised by $\theta$ and $\theta_E$ and hence can be
extended to 3-tori as shown. Since exponentials are a complete and
orthonormal basis on the 3-torus and because we assumed that $s$
and $b$ are band-limited, the $T_{m\,m'\,m''}$ contain all
information about the inverse transform, \eq{eq:totalconv},
$$
T(\phi_E,\phi,\omega) = \hspace{7.5cm}
$$
\begin{equation}
\hspace{2cm}\sum_{m,m',m''=-\lmax}^{\lmax}  T_{m\,m'\,m''} \,
e^{im\phi_E+ im'\phi+im''\omega}.
\end{equation}
Not all tuples $(\phi_E,\phi,\omega)$ correspond to distinct beam
orientations but this redundancy is more than compensated for by
the efficiency of the method.

\section{Discussion}
\label{sec:discussion}
Several remarks about \eq{eq:result} are in
order.
\subsubsection{Computational cost}
Computing the $T_{m\,m'\,m''}$ in \eq{eq:result} costs
\order{ \lmax^4\vert\sin\theta_E\vert  2 \theta/\pi }
operations. The factors $\theta$ and  $\vert\sin\theta_E\vert$ come from the
fact that the
bandlimit $\lmax$ for the sky implies a bandlimit
$\propto 2\theta\lmax/\pi$ on a ring of
radius $\theta$ and hence the ranges of $m$ and $m'$ can be reduced by
factors of $\vert\sin\theta_E\vert$ and $2\theta/\pi$, respectively,
if the rings and the ring of ring centers are not great circles.
Using the Fast Fourier Transform (FFT) algorithm,
the inverse Fourier transform  takes
\order{\lmax^{3}\log \lmax\vert\sin\theta_E\vert  2 \theta/\pi} operations.
 If the
convolved sky is assumed to be real, we have
\begin{equation}
T_{m\,m'\,m''}=T^{\ast}_{-m\,-m'\,-m''},
\end{equation}
reducing memory and processor
requirements by a factor 2.

\subsubsection{Quadrature  and interpolation}

In pixel space each evaluation of $T(\phi_E,\phi,\omega)$ is an
explicit quadrature problem and hence necessarily approximate.
In our
approach, all sums have a finite number of terms and the results
are exact as long as $s$ and $b$ are band-limited.  Quadrature issues
only have to be dealt with if $b$ or 
$s$ are given in pixel space and we have to evaluate the beam and
sky multipole coefficients $b_{lm}$ and $s_{lm}$. The details of which
pixelization to choose on the sphere and how to solve
this generalized quadrature problem for the multipole coefficients are
outside of the scope of this 
work but an
efficient and practical approach to the quadrature problem is implemented in
the {\sc HEALPix} 
package \cite{HEALPix} and will be discussed in
a future publication \cite{Wandeltetal}.

An interesting property of \eq{eq:result} is that as long as
$\lmax$ was chosen appropriately one is guaranteed to have the
convolved sky sampled sufficiently densely for worry-free
interpolation on either of the three indices.

\subsection{Special cases}
We will now discuss certain special cases of \eq{eq:result}.
\subsubsection{Total convolution}
Let us obtain the convolved sky at all possible  beam orientations $\omega$
on an equidistant coordinate grid in $\phi$ (corresponding to the
polar angle) and $\phi_E$ (corresponding to the azimuthal angle). We will refer to this case as the
{\em total convolution}. This can be achieved by evaluating \eq{eq:result} setting
$\theta=\theta_E=\frac{\pi}{2}$. In this case we only need to know
$d^l_{m'm}(\frac{\pi}{2})$. This means we only have to evaluate a single
recursion relation to evaluate the sum on $l$, which simplifies the
computation. The inverse FFT gives the desired result.

A further simplification arises in this case from the fact that if $\theta_E=\frac{\pi}{2}$, not all components
of $T_{m\,m'\,m''}$ are independent.
The redundancy in the parametrization where the polar angle $\phi\in[0, 2\pi)$ leads
to the symmetry
\begin{equation}
T(\phi_E,\phi,\omega)\equiv T(\pi+\phi_E,2\pi-\phi,\pi+\omega).
\end{equation}
This translates into the identity
\begin{equation}
T_{m\,m'\,m''}\equiv(-1)^{m+m''}T_{m\,-m'\,m''},
\end{equation}
which cuts the required memory and computation time by a factor 2.

\subsubsection{Exact or approximate azimuthal symmetry of the beam}
In many practical situations the ``beam'' represents the response function
of an optical system with only mild imperfections. If this is the
case, the beam has only
slowly varying azimuthal structure, implying a
cutoff wavenumber $\mmax$ such that $b_{lm}\sim 0$ for $m\ge \mmax$.
in this case the computational cost for a total convolution scales as
\order{\lmax^3 \mmax\theta \sin\theta_E}.

In the limit of an
azimuthally symmetric beam, $\mmax=0$, we obtain an \order{\lmax^3\theta \sin\theta_E}
method. However, it is known \cite{DriscollHealy} that at least
in principle there exist faster methods for convolution of a
function on the full sphere ($\theta=\theta_E=\pi/2$) with an
azimuthally symmetric  beam which scale as
\order{\lmax^2(\log(\lmax))^2}. We can show how this limit is
obtained from \eq{eq:result} by using the facts
that
the $d^l_{m m'}(\frac{\pi}{2})$ are the Fourier coefficients of the
$d^l_{m m'}(\theta)$ and that $d^l_{m 0}(\theta)=P_{lm}(\theta)$. Then
\eq{eq:result} reduces to the form
\begin{equation}
T(\phi_E,\phi)=\sum_{lm}Y_{lm}(\pi-\phi,\phi_E+\pi/2) b_{l0}s_{lm},
\end{equation}
where the arguments of $Y_{lm}$ are the polar angle and the
azimuthal angle, respectively.
The  algorithm by \cite{DriscollHealy} succeeds precisely in
reducing the computational  cost of evaluating this expression to
\order{\lmax^2(\log^2 \lmax)} under the proviso of the technical
difficulties there outlined.

We note here for completeness, that by choosing a delta function beam
(and hence $b_{lm}=const$), we recover the 
Fourier summation method for the spherical harmonic transform, described in
equations (5.2) to (5.4) in \cite{Risbo}. This computes
\eq{eq:harmonic_expansion} on an equidistant 
coordinate grid by doing Fourier transforms on latitudinal and
longitudinal lines. The forward transform is obtained by simply working all
steps in reverse. 

\subsubsection{Basic scan paths}
\label{sec:basic} Consider an application where the convolution
is required only along a ``basic scan path''.  This is one of the
proposed scan strategies for the  Planck satellite  mission. From our
definition of basic scan paths in section \ref{sec:factor} we see that
they correspond to setting $\omega=0$ in \eq{eq:result}.

Computing the inverse Fourier transform of \eq{eq:result} with
$\omega=0$ just amounts to summing over $m''$. Then only the
two--dimensional Fourier transform
\begin{equation}
T_{m\,m'}(\omega=0)=\sum_{l} s_{lm} d^l_{mm'}(\theta_E)  X_{lm'}
\label{eq:specialresult}
\end{equation}
remains to be evaluated.
The quantity
\begin{equation}
X_{lm}\equiv \sum_{M} d^l_{mM}(\theta)b_{lM}^\ast
\label{eq:precompute}
\end{equation}
can be precomputed. All in all the computational time needed for
evaluating these expressions is ${\cal O}(\lmax^3\theta
\sin\theta_E)$. Storage requirements  scale only as \order{\lmax^2}.

Note that in this case azimuthal symmetry does not
necessarily imply  reduced computational cost. If the beam is
concentrated at the north pole into a region of size $\sigma$ then
$X_{lm}$ will have $m$ modes populated up to $\mmax\sim
{\theta}/{\sigma}$.

Geometrically, the basic scan path corresponds to a 2-torus which
is the section of the 3-sphere of rotations at constant $\omega$.
Note that in this case there is no redundancy in the
parametrization --- every tuple $(\phi_E,\phi,0)$ corresponds to a
distinct beam orientation.

\subsubsection{Perturbations about basic scan paths}

A slight generalization of the previous case are scan paths which are
close to basic but
include a variation in $\theta_E$ from scanning circle to scanning circle. Such paths result for example from precessing
or ``wobbling'' the spin axis of a  scanning satellite.

Such scan paths can be composed by computing several convolutions
along basic scan paths for different angles $\theta_E$  and then
choosing scanning circles at will from among these basic ones.
This method suggests itself if the precession angle is small and
hence a small number of convolutions is sufficient to sample the
variation in $\theta_E$. Convolutions at points which do not
coincide with sampling points can then be determined by
interpolation.

Another approach to this type of problem and further
generalisations are discussed in the next paragraphs.

\subsubsection{Other special cases}
Other potentially interesting special cases of \eq{eq:result} can be
worked out by
fixing any of the parameters to special values and evaluating the inverse
transform, analogous to the calculation for basic
scan paths. For example one obtains expressions for
\begin{itemize}
\item All possible
beam orientations along a circle of constant latitude $\theta_E$. In
this case $\phi$ and $\omega$ have the same meaning; formally,
$T_{m\,m'\,m''}=T_{m\,m'}\delta_{m'\,m''}$ and we obtain an
\order{\lmax^2 \mmax \sin\theta_E} method:
\begin{equation}
T_{m\,m'}=\sum_{l} s_{lm} d^l_{mm'}(\theta_E)b^{\ast}_{l\,m'} .
\end{equation}

\item Individual scanning rings of a basic scan path. Here,
$\omega=0$, and the only free parameter is $\phi$.

\end{itemize}

The details of the calculations for
this and similar cases  are now easy exercises.

\subsubsection{Generalizations}
Further, it is clear from the derivation that  more general types
of paths can be constructed by factorizing the rotation operator
more than twice, so as to generate for example a ring of ring of
rings, etc. For particular applications some of these may be
advantageous, for example if they simplify the interpolation
problem on the output ring set. A specific example is the
precessing scan path mentioned in the previous subsection.
Inserting another rotation operator $\D(0,\theta_P,\phi_P)$
between the two operators in \eq{eq:rotsplit}, and setting
$\omega=0$ produces a set of rings whose centers lie on circles
of radius $\theta_P$ about $theta_E$. This may simplify the
interpolation problem. The rotation corresponding to $\phi_P$ can
be sampled sparsely if $\theta_P$ is small ($\mmax_P\sim
\sin\theta_P\sin\theta_E\lmax$ with obvious notation) and the
interpolation problem becomes simpler.

\section{Conclusions}

This paper presents a general algorithm which greatly reduces the
computational cost of convolving two bandlimited but otherwise
arbitrary functions on the sphere. The speedup increases linearly
with the smallest angular scale of the smoother of the two
functions in the problem. The scalings of the necessary operation
counts are discussed in detail in section five.

We quote in the appendix formulas showing how the ideas presented
in this paper can be applied to the inverse problem of
``deconvolution'' by speeding up the iterative solution of the
normal equation in an analogous way.

This paper focuses on the
convolution of scalar valued functions on the sphere such as
temperature, elevation, etc. In order to be able to deal with
the polarization of the cosmic microwave background we extend the
methods presented here to tensor valued functions on the sphere in
reference \cite{CMFAWG}.

The algorithms which are presented here are already being used as
a core component of the prototype simulation pipeline of the
Planck  satellite. To give an example for the timing gains one
makes by applying this  method, we computed the following case:
both sky and beam were interpolated and pixelised very densely,
with millions of pixels each to resolve the steep variations over
many orders of magnitude. The bandlimit was somewhat generously
chosen as $\lmax=1024$. Then the convolution of the sky with the
beam of a single detector for a whole year of mission data,
consisting of $(2049)^2\sim 4\times10^6$ convolved samples along a basic scan
path was generated in less than 15 minutes on a single Silicon
Graphics R10000 processor. This compares with several days of
computation on a severely coarsened sampling grid with several hundred
times fewer samples on the same machine, using the adaptive mesh
method \cite{B99}. For the same resolution which we achieved
with our methods, the adaptive mesh code would have run for months.

Due to our methods, future CMB missions can go beyond having to
approximate the treatment of realistic beams. Our methods lend
themselves to being used in conjunction with iterative map--making
methods  to remove from the data  artefacts which are due to beam
distortions and far side lobes (see appendix).

Lastly, we feel that the geometric constructions, analogies to group
properties and algebraic results we introduce in this article may be
useful more generally for CMB data analysis and plan to explore these
issues in future work.  

\begin{acknowledgments}
We would like to thank F.\ Hansen for coding help in the early
stages of this project, F.\ Bouchet for suggesting consideration
of the transpose operation and R.\ Caldwell for reading the
manuscript before submission. Part of this work was supported by
the Dansk Grundforskningsfond through its funding for TAC. BDW is
supported by the NASA MAP/MIDEX program.
\end{acknowledgments}

\appendix

\newcommand{\eg}{{\it e.g.\ }}
\newcommand{\R}{{\sf I\!R}}
\newcommand{\id}{{\mathbf{\rm 1\hspace{-.105cm}I}}}
\newcommand{\ANA}{\mathbf{A^TA}}
\newcommand{\AN}{\mathbf{A^T}}

\section{The convolution transpose}

We sketch here how to set up the inverse problem of reconstructing
the true sky from the convolved observations.
If we start with a noise free set of convolutions, then the equation to
be inverted in order to estimate the true underlying sky is, schematically,
\begin{equation}
\mathbf{A\, s}  =\mathbf{d} \;.
\label{normal}
\end{equation}
Here, $\mathbf{s}$ is the true sky, $\mathbf{d}$ is the vector
containing the time-ordered data after observation. The convolution operator is
represented by $\mathbf{A}$.

The least-square estimator for the true sky, $\mathbf{\hat{s}}$ then
satisfies the {\em normal equation}
\begin{equation}
\mathbf{A^TA\, \hat{s} }  =\mathbf{A^T\,d} \;.
\label{eq:estimator}
\end{equation}
For a perfect observation with a delta function beam,
$\ANA\equiv \id$. So it may be
reasonable to expect that we can make progress by considering a mildly
imperfect optical system and consider iterative techniques for solving
the normal equation. In this
case the ability to solve for $\hat{s}$ iteratively (e.g.~using a Conjugate
Gradient technique) relies on convergence (which is assured up to
numerical effects because the normal matrix $\ANA$ is positive
definite and being able to compute the matrix products in (\ref{eq:estimator})
quickly. The application of $\mathbf{A}$ can be computed efficiently
using the formulae set out in sections four and five. We now
find an algorithm for the efficient application  of
$\mathbf{A^T}$, the {\em transpose convolution}.

\subsection{Applying the transpose convolution}

We can write down the expression for $\mathbf{A^T}$ in a similar way to
\eq{eq:totalconv}
$$y(\vec{\gamma})=\hspace{8cm}$$
\begin{equation}
\int d\phi_E d\phi d\omega
\left\lbrack\D(\phi_E,\theta_E,0)\D(\phi,\theta,\omega)b\right\rbrack\!\!(\vec{\gamma})^\ast
\,T(\phi_E,\phi,\omega).
\label{eq:atrans}
\end{equation}
Now the derivation is analogous to the one preceding \eq{eq:result},
and $y(\vec{\gamma})$ is given in terms of the Wigner $d$ functions as
\begin{equation}
y(\vec{\gamma})=\sum_{lmm'm''}Y^\ast_{lm}(\vec{\gamma})d^{l}_{mm'}(\theta_E)d^{l}_{m'm''}(\theta_E)b^\ast_{lm''} T_{mm'm''}
\label{eq:atres}
\end{equation}

This formula can be generalised or applied to special cases just
as we showed in section \ref{sec:discussion} for \eq{eq:result}.


\begin{thebibliography}{10}

\bibitem{review}
W. Hu, N. Sugiyama, and J. Silk, Nature {\bf 386}, 37 (1997)

\bibitem{borrill}
J. Borrill, Phys. Rev. {\bf D59},  027302  (1999).

\bibitem{BCJK}
J.~R. Bond, R.~G. Crittenden, A.~H. Jaffe, and L. Knox, Computing in Science
  and Engineering {\bf 1},    (1999).

\bibitem{haslam}
C.~G.~T. Haslam, A\&AS {\bf 47},  1  (1982).

\bibitem{SFD98}
D. Schlegel, D. Finkbeiner, and M. Davis, Astrophys.\ J.\ {\bf 500},  525
  (1998).

\bibitem{WS95}
M. White and M. Srednicki, Astrophys.\ J.\ {\bf 443},  6  (1995).

\bibitem{K99}
L. Knox, \prd {\bf 60},  103516  (1999).

\bibitem{Wu00}
 J. H. P. Wu {\it et~al.},  astro-ph/0007212 (unpublished).

\bibitem{dM98}
P. de~Maagt, A.~M. Polegre, and G. Crone, {\em Straylight evaluation of the
  Carrier Configuration}, ESA technical report, (1998)

\bibitem{D98}
J. Delabrouille, {\em PhD thesis}, 1999.

\bibitem{B99}
C. Burigana {\it et~al.},  astro-ph/0010113, (unpublished)

\bibitem{CMFAWG}
A. Challinor {\it et~al.},  astro-ph/0008228, Phys.~Rev.~D in press,

\bibitem{BrinkSatchler}
D.~M. Brink and G.~R. Satchler, {\em Angular Momentum} (Clarendon Press,
  Oxford, 1975).

\bibitem{Risbo}
T. {Risbo}, Journal of Geodesy {\bf 70},  383  (1996).

\bibitem{DriscollHealy}
J.~R. Driscoll and D.~M. Healy, Adv.\ in Appl.\ Math. {\bf 15},  202  (1994).

\bibitem{HEALPix} 
The HEALPix Homepage,\\
{\tt http://www.eso.org/$\sim$kgorski/healpix/}

\bibitem{Wandeltetal}
B.~D. Wandelt {\it et~al.}, {\em Fast Multipole Quadrature on the
Sphere with Applications to HEALPix}, in preparation (unpublished).

\end{thebibliography}

\end{document}